\pdfoutput=1

\documentclass[11pt]{article}

\usepackage[final]{acl}

\usepackage{times}
\usepackage{latexsym}

\usepackage[T1]{fontenc}
\usepackage{CJKutf8}

\usepackage[utf8]{inputenc}

\usepackage{microtype}

\usepackage{inconsolata}

\usepackage{graphicx}
\usepackage{booktabs} 
\usepackage{tabularx}
\usepackage{multirow}
\usepackage{adjustbox}
\usepackage{amsmath}
\usepackage{wasysym}
\usepackage{amssymb}
\usepackage{tcolorbox}
\usepackage{listings}
\usepackage{needspace}

\newif\ifproofread

\newcommand{\update}[1]{%
\ifproofread
\textcolor{purple}{#1}%
\else
#1%
\fi
}

\title{SPOT: Bridging Natural Language and Geospatial Search for\\Investigative
Journalists}

\author{
    Lynn Khellaf\thanks{Equal Contribution} \quad Ipek Baris Schlicht\footnotemark[1] \quad Tilman Mira\ss{} \\
    \textbf{Julia Bayer} \quad \textbf{Tilman Wagner} \quad \textbf{Ruben Bouwmeester} \\
    \update{Deutsche Welle} Innovation, Bonn/Berlin \\ 
    \url{https://innovation.dw.com/}\\
    \texttt{hey@findthatspot.io}
}

\newcommand{\xmark}{\color{red}{\text{\sffamily X}}}
\newcommand{\jp}[1]{\begin{CJK}{UTF8}{min}#1\end{CJK}}

\proofreadfalse

\begin{document}
\maketitle
\begin{abstract}
OpenStreetMap (OSM) is a vital resource for investigative journalists \update{doing} geolocation verification. However, \update{existing tools to query OSM data} such as Overpass Turbo require familiarity with complex query languages, creating barriers for non-technical users. We present SPOT, an open source natural language interface that makes OSM's rich, tag-based geographic data more accessible through intuitive scene descriptions. SPOT interprets user inputs as structured representations of \update{geospatial} object configurations using fine-tuned Large Language Models (LLMs), with results being displayed in an interactive map interface. While more general geospatial search tasks are conceivable, SPOT is specifically designed for use in investigative journalism, addressing real-world challenges such as hallucinations in model output, inconsistencies in OSM tagging, and the noisy nature of user input. It combines a novel synthetic data pipeline with a semantic bundling system to enable robust, accurate query generation. To our knowledge, SPOT is the first system to achieve reliable natural language access to OSM data at this level of accuracy. By lowering the technical barrier to \update{geolocation} verification, SPOT contributes a practical tool to the broader efforts to support fact-checking and combat disinformation.
\end{abstract}

\section{Introduction}
Investigative journalists frequently rely on OpenStreetMap (OSM)~\cite{osm} as a vital tool for geolocation verification or research because of its detailed and comprehensive coverage of various locations. However, non-technical users face challenges due to required knowledge of query languages (such as OverpassQL\footnote{\url{https://wiki.openstreetmap.org/wiki/Overpass_API/Overpass_QL}}) for data retrieval.

Although language models have been applied to relational database interactions, their use in OSM-based applications is still limited and not tailored to the needs of investigative journalists. \citet{heidelberg} and \citet{simon} for instance introduced datasets and applications that employ neural-network-based semantic parsers to transform natural language into intermediate query formats. Similarly, \citet{staniek-etal-2024-text} introduced the OverpassT5 model along with benchmarking data for directly querying OSM. However, prior datasets are not directly applicable to the current use case, as they assume prerequisite knowledge of OSM functionalities. While there are AI-powered geolocation tools available to support investigative journalists, they either don't or fail to work effectively with unstructured text inputs~\cite{earthkit, geospy}, or are based on source code that is not publicly available or utilize closed Large Language Models (LLMs)~\cite{geoguessr}.

\begin{figure*}[!ht]
    \centering 
    \includegraphics[width=\linewidth]{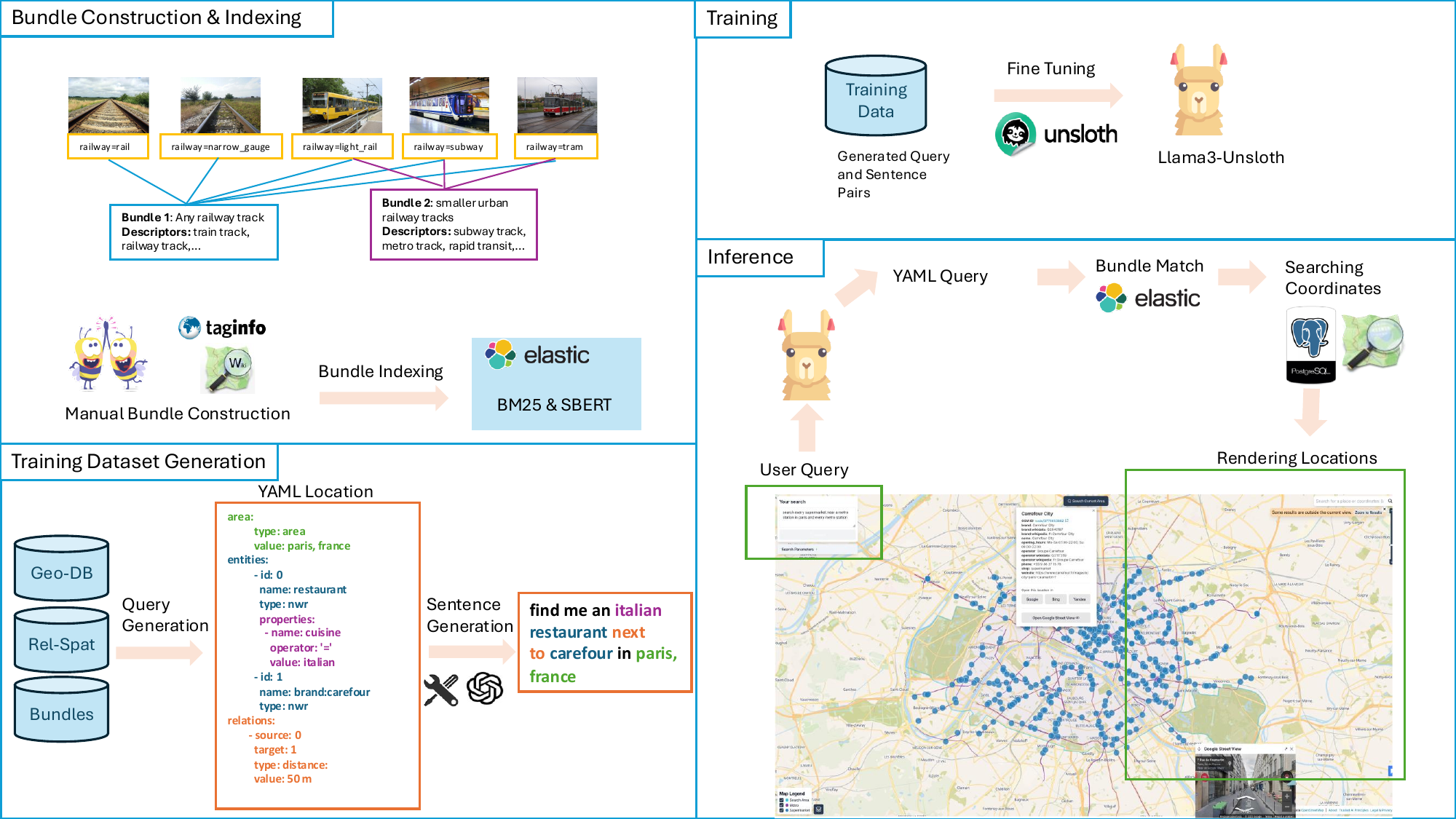} 
    \caption{Overview of SPOT's OSM-based pipeline, from tag bundle indexing and semantic search, through artificial sentence and YAML pair generation, to model fine-tuning and interactive inference.} 
    \label{fig:architecture} 

\end{figure*}

To this extent, we present SPOT, an AI-powered, fully open source and open weight geospatial tool designed for investigative journalism, although other potential applications are conceivable. As illustrated in Figure~\ref{fig:architecture}, SPOT includes a pipeline for generating artificial training data tailored to user requirements and the OSM tagging system. Its backbone model leverages LLaMA 3~\cite{llama}, which is fine-tuned on the generated data. During inference, SPOT transforms user input into YAML-based queries which are enriched with predefined OSM tag bundles by using a semantic search engine. Additionally, SPOT provides a user-friendly graphical interface that enables users to seamlessly enter their unstructured search requests, with results displayed interactively on a map. Places of interest can \update{be further} explored \update{in detail} via integrated external tools such as GoogleStreetView. SPOT is publicly accessible at \url{https://www.findthatspot.io/}, with its source code hosted on GitHub\footnote{Source code: \url{https://github.com/dw-innovation/kid2-spot}}. Moreover, the fine-tuned LLaMA 3 model, along with other benchmarked LLMs (detailed in Section~\ref{sec:experiments}), is available on HuggingFace\footnote{Model weights: \url{https://huggingface.co/DW-ReCo}}.

\section{Related Work}
\begin{table*}[ht]
\centering
\adjustbox{width=\linewidth}{
\begin{tabular}{lllll}
\toprule
\textbf{Tool} & \textbf{Input} & \textbf{Customization} & \textbf{External Data Integration} & \textbf{Open Source}\\ 
Overpass Turbo~\cite{ot} & OT Query & via Query & \checkmark & \checkmark \\ 
GeoGuessr GPT~\cite{geoguessr} & Unstructured Text & via Chat & \xmark & \xmark \\
GeoSpy~\cite{geospy} & Image & NA & \checkmark & \xmark \\ 
EarthKit~\cite{earthkit} & Semi-structured Text & via Query & \checkmark & \checkmark \\ 
SPOT & Unstructured Text & User Guided Search & \checkmark & \checkmark \\
\bottomrule
\end{tabular}}
\caption{Comparison of OSM-based, AI-supported geolocation verification tools.}
\label{tab:geolocation-tools}
\end{table*}
\subsection{Text-to-Structured Language}
Several research studies have explored ways for users to interact with databases without requiring technical knowledge of structured query languages. The most common approach is to transform natural language questions into SQL queries (text-to-SQL) to facilitate interaction with relational databases\update{, which is closely related to the current use case}. \update{Recent advances in this area have explored both prompt-based methods and parameter-efficient tuning of LLMs~\cite{zhu2024large,shi2024survey}. For example, \citet{jang2023exploring} applied adapter tuning to T5~\cite{t5}, while \citet{DBLP:conf/sigmod/ZhangMFMG0LL24} used adapter tuning and merging on LLaMA. Other work has focused on prompt engineering: \citet{DBLP:journals/pvldb/GaoWLSQDZ24} proposed DAIL-SQL to improve example selection in in-context learning, and \citet{lee-etal-2025-mcs} introduced MCS-SQL, which uses multi-prompting for text-to-SQL generation.} 

Despite the growing importance of OSM for applications such as geo-verification in journalism, natural language interaction with OSM has been relatively under-researched compared to text-to-SQL. Some research~\cite{heidelberg,simon} proposes the use of semantic parsers to convert natural language queries into intermediate representations that include elements from OSM tags, which can be used to create downstream OSM queries. In contrast, \citet{staniek-etal-2024-text} tackled the direct text-to-OverpassQL task, creating a dataset of natural language inputs paired with their corresponding \update{OverpassQL} queries. They also introduced a task-specific evaluation metric that considers surface string similarity, semantics, and syntax. Their evaluation indicated that explicit pre-training of sequence-to-sequence models like OverpassT5 was not beneficial, while few-shot prompting with GPT-4 performed the best.

Unlike previous approaches, the intermediate representation step in SPOT is multi-layered. To handle variations in query styles (e.g., typos or different terms for the same object) and to allow for updates to OSM tags without needing to retrain the language model, we employ multiple processing steps. SPOT queries are structured in YAML and initially do not contain any OSM tag elements. In a second step, object and property names are passed through a semantic search engine and replaced with the best-fitting OSM tag bundles required for the final OSM database request. We fine-tuned an instance of LLaMA 3 to generate the initial YAML. This state-of-the-art LLM is vastly more performant than our earlier T5-based approach~\cite{khellaf2023spot}, in which we encountered limitations addressing several key requirements.

\subsection{OSM Datasets} 

The datasets~\cite{heidelberg,simon,staniek-etal-2024-text} are currently the only publicly available resource designed for natural language interaction with OSM. They allow users to query OSM using its tagging system, based on coordinates, specific tag types or meta-information such as changes made by particular users. These datasets, however, are primarily intended for users who are familiar with OSM's tagging logic, making them difficult to use for those without prior experience.

In contrast, our tool is designed for visual location verification, allowing \update{users} to perform the search using natural language descriptions without requiring OSM expertise. Our approach focuses on visual features such as objects, their properties and the spatial relationships between them, while excluding meta-information irrelevant to the task. For this purpose, we have developed a pipeline for artificial data generation tailored to these specific needs.

\subsection{Comparison of Geolocation Tools}
There are numerous geolocation tools that have a similar target audience, with and without AI support. Among the most popular for investigative journalists are the original Overpass Turbo~\cite{ot} (not using AI), GeoGuessr GPT~\cite{geoguessr}, GeoSpy~\cite{geospy} and EarthKit~\cite{earthkit}. Table~\ref{tab:geolocation-tools} contains a design comparison of the aformentioned tools with SPOT. Both SPOT and GeoGuessr GPT (which uses ChatGPT with a custom prompt) accept unstructured text as input, while the other tools rely on structured queries, images, or semi-structured text. In the case of EarthKit, users are presented with OSM tags and must manually select the relevant ones to complete their query.

Of these tools, only SPOT and EarthKit offer full stack open source software and AI models, allowing anyone to host \update{them} on their own infrastructure. In terms of integration, GeoGuessr GPT does not connect to any external tools or OSM other than GPT, while EarthKit only integrates with OSM. The remaining tools offer integration with Google Maps or Google Street View. In addition to linking to the location on Google, Bing and Yandex, SPOT also \update{features an OpenStreetView.com integration for a detailed view of} identified locations, increasing its utility for investigative work.  

\section{Overview of SPOT} 
As shown in Figure~\ref{fig:architecture}, SPOT has four main components: bundle construction and indexing, training data generation, training and inference. Each component is briefly described in the following subsections.

\subsection{Bundle Construction and Indexing} \label{sec:bundles}

To bridge the gap between natural language and the OSM tagging system, we developed a static \textit{bundle list} that groups visually similar (individual or combinations of) OSM tags. This list maps natural language descriptors to relevant OSM tags, taking into account the ambiguity and variability of everyday language. For example, terms such as \textit{light rail}, \textit{subway} and \textit{tram} are all mapped to the same bundle representing ``smaller urban railway tracks''. This approach helps to mitigate inconsistencies in OSM tagging, where multiple tags or tag combinations can refer to objects that are frequently referred to by the same terms.

To make them searchable, the bundle lists are indexed via Elasticsearch\footnote{\url{https://www.elastic.co/elasticsearch}}. We index both the raw text and its semantic embeddings to deal with typos and paraphrases. The semantic embeddings are \update{vectorized} using the \texttt{all-MiniLM-L6-v2} version of the SBERT sentence transformer~\cite{reimers2019sentence}. This setup allows for a hybrid search approach that combines BM25 with SBERT-based retrieval.

\subsection{Training Dataset Generation} \label{sec:dataset_generation} 

\begin{figure}[!ht]
    \centering 
    \includegraphics[width=\linewidth]{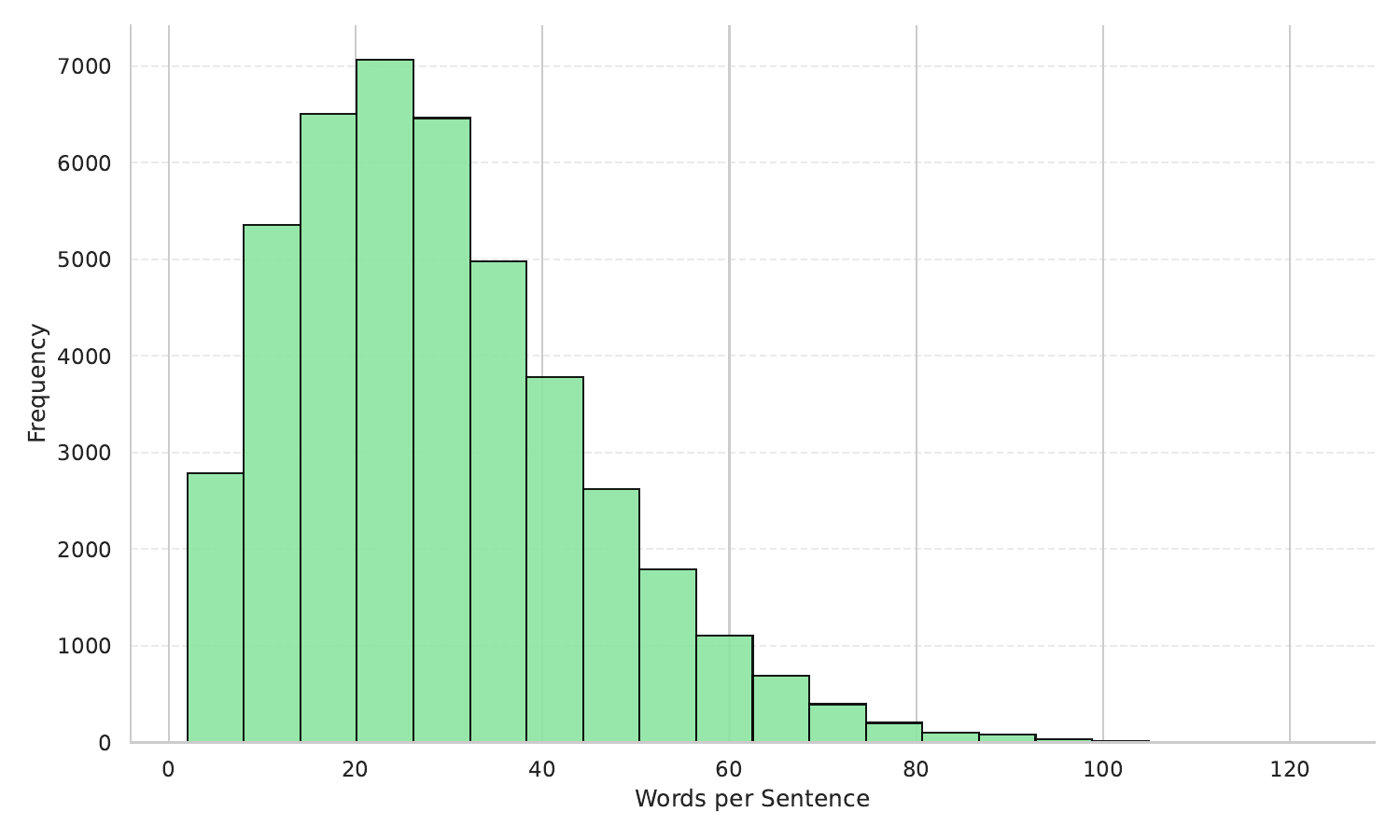} 
    \caption{\update{Sentence length distribution of the generated sentences}} 
    \label{fig:sent_length} 

\end{figure} 

\begin{figure}[!ht]
    \centering 
    \includegraphics[width=\linewidth]{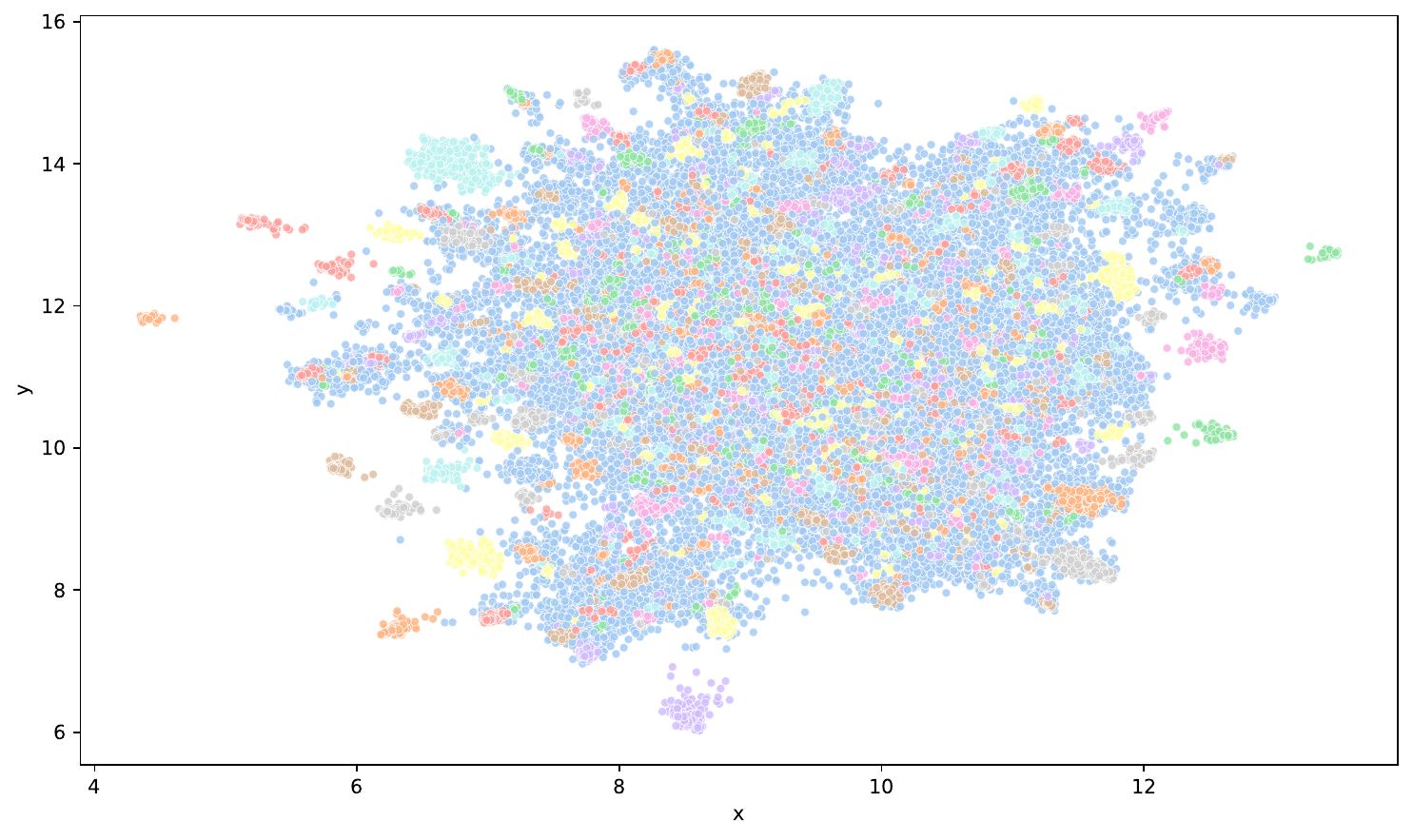} 
    \caption{\update{Semantic Diversity Visualization of sentence embeddings of the generated sentences using UMAP and HBSCAN. Blue dots indicate the noisy points that do not belong to any clusters (16,416 points in total).}} 
    \label{fig:data_divers} 

\end{figure} 

\begin{table*}[ht]
\small
\centering
\begin{tabularx}{\textwidth}{X}
\toprule
- Find a tattoo shop and a doityourself shop, both within 2.5 ft of each other. \\
- Find a restroom and an american football field in \jp{米林根, 巴登-符腾堡, 德国}, no more than 28 meters apart. \\
-In the region of Ward County, North Dakota, United States, seek out a campsite alongside a production studio, specifically one that is situated on a street whose name concludes with the suffix "-der-Tann-Straße." \\
- Let's see. I'm looking for a \jp{新星堂}. Then there's a moving walkway. It has a traffic lane numbered 484 and a car lane numbered 581. I also need to find a monument whose name starts with ""emin du Ro"". All of these should be found within a distance of 75556 miles from one another. \\
- Could you kindly locate a play area within the confines of Comuna Vadu Moţilor? \\
- Find a bowling cemter located three hundrd kilomters away from a camera shop. \\
\bottomrule 
\end{tabularx}
\caption{\update{Examples from the training dataset showing different features (e.g. long/short sentence, properties, typos, non-Latin alphabet, etc.).}}
\label{examples_training}
\end{table*}

Prior to development, we conducted a user study with the in-house SPOT development team \update{and our expert OSINT community} to collect descriptions of scenes based on images. From this study, we derived a list of user requirements (Appendix \ref{appendix:requirements}) to guide system development. Key findings included the high prevalence of generic terms for objects and spatial relations, as well as frequent typos and grammar errors. 

As illustrated in Figure~\ref{fig:architecture}, we designed a novel YAML-based structure to simplify data handling, overcoming the challenges associated with JSON's strict syntax~\cite{yamljson}. The structure contains all relevant information, namely search area, entities, properties, and spatial relations. We implemented a framework that creates any number of YAML combinations via random draft of values for the semantic fields. Relation types distinguish between distance and contains relations, as inspired by the user requirements. In addition to specific distance values (such as \textit{within 100 meters}), the model is trained to translate vague relative spatial terms (such as \textit{nearby}, \textit{next to}) into concrete values (\textit{next to} for instance is defined as 50 meters, the full list in Appendix \ref{appendix:relspat}). The multi-lingual area names used in the artificial data are extracted from the public map database NaturalEarthData\footnote{\url{https://www.naturalearthdata.com/}}. 
The information from the YAML queries with additional text style (e.g. typos) and persona (e.g. fact-checker) specifications is then used to dynamically generate prompts, which is in turn used to turn the YAML into a synthetic natural query sentences using GPT-4o~\cite{gpt4}. 

\update{ In total, we used 7 personas and 5 writing styles, we provide them in Appendix~\ref{appendix:persona_style}.} The number of generated samples for training is 43976, 2350 of which form the development set. An example prompt is shown in \update{Table}~\ref{tab:app22}. \update{As shown in Figure~\ref{fig:sent_length}, the generated dataset contains different length of sentences. To evaluate the semantic diversity of the generated dataset, we first performed sentence embedding using SBERT. We then used UMAP~\cite{mcinnes2018umap-software} to project these high-dimensional embeddings into 2D space for visualization, while preserving local semantic relationships. UMAP was configured with 50 nearest neighbours, a minimum distance of 0.1, a target dimensionality of two, and a fixed random seed to ensure reproducibility. We then applied HDBSCAN~\cite{10.1145/2733381} to the resulting 2D embeddings. HDBSCAN is a density-based clustering algorithm that can detect clusters of varying shapes and identify outliers. HDBSCAN was configured with a minimum cluster size and minimum samples parameter both set to 5. The algorithm identified 1,274 distinct clusters but did not assign cluster labels to 16,416 sentences, treating them as noise. A graph of the result can be seen in Figure~\ref{fig:data_divers}. The considerable number of clusters, along with a substantial proportion of unclustered sentences, indicates that the generated dataset exhibits significant semantic diversity.}

\subsection{Training and Inference} \label{training}
We fine-tuned an open-source LLM on the synthetic dataset (described in Section~\ref{sec:dataset_generation}) by using the \texttt{unsloth library}\footnote{\url{https://unsloth.ai/}}. The fine-tuning process employed Low-Rank Adaptation~\cite{hulora} with a rank of 32 and an alpha scaling factor of 64. Training was conducted with a batch size of 8 and the learning rate was set to 1e-5 with a weight decay of 0.01. Early stopping was activated with a patience of 10 epochs and evaluation was performed every 200 steps.


We host the SPOT language model using HuggingFace Inference Endpoints\footnote{\url{https://ui.endpoints.huggingface.co/}}. A backend built with FastAPI\footnote{\url{https://fastapi.tiangolo.com/}} handles post-processing of the model output, such as replacing names with corresponding OSM tags. The backend forwards user queries to a PostgreSQL database with the PostGIS extension, indexed with the OSM planetary dataset\footnote{\url{https://wiki.openstreetmap.org/wiki/Planet.osm}}, to retrieve spatial coordinates and details about the detected objects. The results are then finally displayed on an interactive map in the UI.

\section{Experiments} \label{sec:experiments}
\begin{table}[!ht]
\small
    \centering
    \begin{tabular}{lc}
    \toprule
         \textbf{Total} & 195 samples \\
     \midrule    
         \textbf{Named area} & 143 samples\\
         \textbf{No Area (bbox)} & 52 samples\\
         \textbf{Properties} & 63 samples\\
         \textbf{Typos} & 36 samples\\  
         \textbf{Grammar Mistakes} & 39 samples\\  
         \textbf{Relative Spatial Terms} & 43 samples\\
         \textbf{Contains Relation} & 48 samples\\
         \textbf{Distance Relation} & 121 samples\\
    \bottomrule
    \end{tabular}
    \caption{Breakdown of samples in the benchmarking dataset.}
    \label{tab:benchmarking_dataset}
\end{table}

\begin{table*}[ht]
\small
\centering
\begin{tabularx}{\textwidth}{X}
\toprule
- all Don Quijote that are in a retail building with a purple roof coluor in \jp{東京都} \\
- Find me a bus platform next to a Cheesecake Factory restaurant and a building with a
red roof in Dubrovnik. \\
- Focus on Arch, Switzerland. Find a restaurant within 1.5 km of a bus station. The
restaurant should have a public toilet inside. \\
- Search for a planetarium containing a public toilet. It should be within 85,800 yards of a
public clock. \\
- Find a speet kamera within 100 meater from antenna in Paraiba \\
- I'm looking for a supermarket from a brand ending in "ermarché" with a parking lot next to it and a power line running past it in less than 15 meters distance. \\
\bottomrule
\end{tabularx}
\caption{\update{Examples from the benchmarking dataset.}}
\label{tab:benchmarking_examples}
\end{table*}

\begin{table*}[ht!]
    \centering
    \small
    \begin{tabular}{lll}
    \toprule
        \textbf{LLM} & \textbf{Company} & \textbf{Unsloth's Version} \\
        \midrule
        Mistral & Mistral & \texttt{unsloth/Mistral-Nemo-Base-2407-bnb-4bit} \\
        LLaMA 3 & Meta & \texttt{unsloth/llama-3-8b-bnb-4bit} \\
        Phi & Microsoft & \texttt{unsloth/Phi-3-medium-4k-instruct-bnb-4bit} \\
        Qwen2.5 & Alibaba & \texttt{unsloth/Qwen2.5-14B} \\
\bottomrule
    \end{tabular}
    \caption{Open source LLMs that were examined as potential semantic parsers with their company name and model code from Unsloth~\cite{unsloth}.}
    \label{tab:summ}
\end{table*}

\subsection{Experimental Setup}
\noindent
\textbf{Benchmarking Dataset.}
We constructed a benchmarking dataset consisting of real user queries to assess the viability of several candidate LLMs as query translators. The queries were generated by a pool of investigative journalists, fact-checkers, and verification experts from Deutsche Welle while trying to geolocate sample images using an early version of SPOT. The resulting list was then filtered based on how well the queries aligned with the OSM database structure and its resulting limitations. Table~\ref{tab:benchmarking_dataset} shows statistics on the prevalence of different requirements in the dataset.
Table~\ref{tab:benchmarking_examples} highlights some example queries from this study. These sentences showcase some aspects of the linguistic variety the system might be faced with and needs to handle.

\noindent
\textbf{Evaluation Metric.}
As evaluation metric, we evaluated the percentage of the matches across areas, entities, properties and relations. Since the entity and property names detected by the model might be correct but not covered by the static bundle list, we employed the SBERT transformer also used for the bundle indexing. We considered a ground truth and a model prediction a match if their cosine similarity exceeded 0.8. We additionally counted the number of hallucinated/omitted entities and properties.

\subsection{Results}
\begin{table*}[!ht]
    \centering
\small
    \begin{tabular}{lllllll}
    \toprule
    \textbf{Adaptation} & \textbf{Model} & \textbf{Area} & \textbf{Entity} & \textbf{Entity*} & \textbf{Property} & \textbf{Relation}\\
    \midrule
     \textbf{Zero-shot} & \multirow{2}*{\textbf{GPT-4o}} & 88.14 & 2.28 & 90.21 & 3.03 & 9.8 \\
    \textbf{One-shot}  & & 89.18 & 1.13 & 92.03 & 10.96 & 11.11 \\
    \midrule
    \multirow{4}*{\textbf{Adapter Tuning}} & \textbf{mT5} & 88.21 & 72.34 & 90.02 & 48.89 & 37.01\\
    & \textbf{Mistral} & \textbf{93.33} & 82.54 & 95.01 & \textbf{56.58} & 45.45 \\   
    & \textbf{Phi}  & 92.82 & 79.59 & 94.10 & 53.33 & \textbf{53.90}\\   
    & \textbf{LLaMA 3} & 92.31 & \textbf{81.41} & 96.15 & 50.00 & 48.05 \\   
    & \textbf{Qwen2.5} & 92.31 & \textbf{82.31} & \textbf{95.69} & 51.95 & 52.60 \\ 
    \bottomrule
    \end{tabular}
    \caption{Accuracy of the models in identifying areas, entities, properties and relations. Entity* is the accuracy when associated properties \update{are excluded}. \textbf{Bold results} are the top results.}
    \label{tab:res_1}
\end{table*}

\begin{table}[!ht]
    \centering
\adjustbox{width=\linewidth}{
    \begin{tabular}{llllll}
    \toprule
    \textbf{Adaptation} & \textbf{Model} & \multicolumn{2}{c}{\textbf{Entity}} & \multicolumn{2}{c}{\textbf{Property}} \\
    & & \textbf{Missed} & \textbf{Hallucinated} & \textbf{Missed} & \textbf{Hallucinated} \\
    \midrule
     \textbf{Zero-shot} & \multirow{2}*{\textbf{GPT-4o}} & 48 & 37 & 53 & 10 \\
    \textbf{One-shot}  & & 40 & 34 & 50 & 11  \\
    \midrule
    \multirow{4}*{\textbf{Adapter Tuning}} & \textbf{mT5} & 51 & 31 & 15 & 6 \\
    & \textbf{Mistral} & 27 & 21 & 17 & 6 \\   
    & \textbf{Phi} & 30 & 22 & 18 & 7 \\   
    & \textbf{LLaMA 3} & 20 & 16 & 18 & 7 \\   
    & \textbf{Qwen2.5} & 23 & 17 & 19 & 6 \\ 
    \bottomrule
    \end{tabular}}
    \caption{The number of omitted/hallucinated entities and properties of each tested model. }
    \label{tab:res_2}
\end{table}

\begin{figure*}[!ht]
    \centering
    \includegraphics[width=\linewidth]{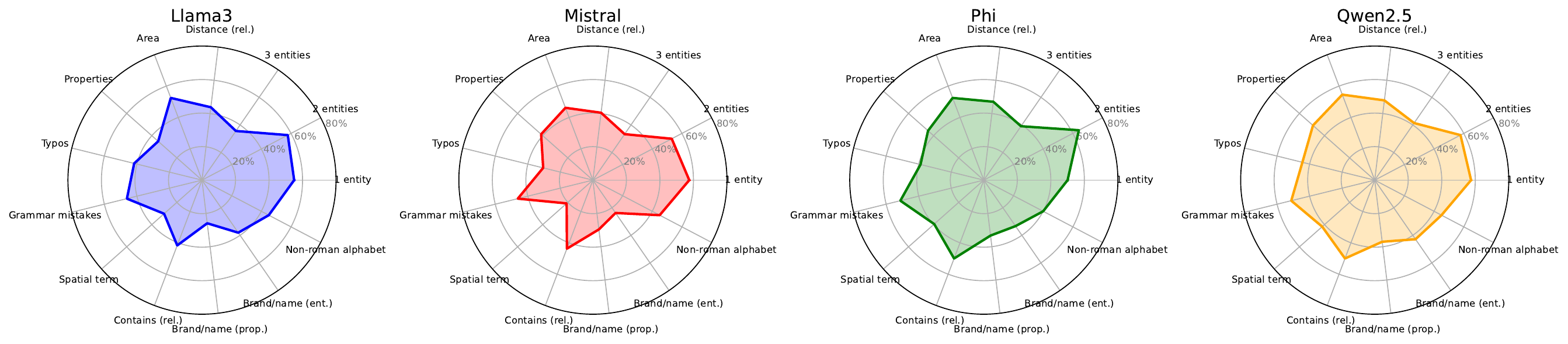}
    \caption{Analysis of LLaMA 3, Mistral, and Phi regarding the ratio of perfect YAML generations various metadata categories. It highlights inter- and intra-model differences in feature handling.}
    \label{fig:finegrained}
\end{figure*}
We evaluated several LLMs as semantic parsers. As a baseline, we used the multilingual T5 variant, mT5, which has shown strong performance in past studies on the generation of structured output despite its relative small size~\cite{khellaf2023spot,staniek-etal-2024-text}. To adapt mT5 to our task, we applied LoRa adapter learning. In addition, we obtained baseline results from GPT-4o by testing it with zero-shot and few-shot prompting (the full prompts are provided in Appendix~\ref{appendix:prompting}).

We then compared the baseline results with several widely used open LLMs from different companies: LLaMA 3~\cite{DBLP:journals/corr/abs-2407-21783} from Meta, Mistral~\cite{jiang2023mistral} from Mistral AI, Phi~\cite{abdin2024phi} from Microsoft, and Qwen~\cite{qwen25} from Alibaba. We applied adapter training as detailed in Section~\ref{training} to the quantized versions of their (due to hardware constraints) small/medium models (as \update{summarized} in Table~\ref{tab:summ}).

As shown in Figure~\ref{tab:res_1}, the fine-tuned LLMs outperformed both GPT-4o and mT5 in all aspects. All fine-tuned LLMs have similar scores for areas, entities, and entities without properties. Noticeably high scores were achieved by Mistral for property, and Qwen2.5 for relation prediction. Qwen2.5 having the most parameters could indicate that relation identification is a task that requires advanced reasoning skills. Furthermore, the fine-tuned LLMs generated fewer hallucinations and omissions compared to the baseline models (shown in Table~\ref{tab:res_2}).

We performed a more nuanced analysis of the generated outputs using meta tags, indicating the use of area names, properties, typos, grammar mistakes, spatial terms, brand names (as entities or properties), non-Roman characters, the presence of distance or contains relations, and the number of entities up to three. The percentage of perfectly generated YAML queries for each category is shown in Figure~\ref{fig:finegrained}. Faulty grammar, typos, and non-Roman characters in particular posed a challenge to the models. Despite these similarities, some model-specific differences are visible, such as Phi and Qwen2.5 performing slightly better when relations were defined using spatial terms.

Finally, we assessed whether the generated output was parsable, as a well-formatted output is essential for the rest of the query pipeline. Based on our benchmark data set, only LLaMA 3 and GPT-4o consistently produced parsable output, leading to the selection of LLaMA 3 as the primary parser for SPOT. A custom parser was deemed too unreliable and potentially detrimental to the inference speed. Although not specifically fine-tuned in languages other than English, the model appears to be able to interpret queries in a variety of languages, although this was not further tested.

\section{Conclusion}

SPOT represents a significant step forward in making OSM more accessible to non-technical users, particularly investigative journalists, through an easy-to-use natural language interface. By addressing the complexity of OSM query languages with a data pipeline that generates any amount of synthetic data, a static list of descriptors, and tag bundles that allow users to perform geospatial searches using their natural language, SPOT improves the usability of OSM data. Our evaluations demonstrate its ability to handle different linguistic styles, grammatical errors and different types of object relationships, achieving state-of-the-art performance in query interpretation with fine-tuned LLaMA 3 and other LLMs. This work bridges the gap between complex geospatial query languages and practical, intuitive interfaces.

Despite its strengths, SPOT's reliance on synthetic data, limits in hardware and a small benchmark dataset highlight potential avenues for future improvement. We further aim to expand language support, add multimodal features such as image queries, and explore an alternative chat interface to further improve usability. \update{Lastly, we plan to conduct comprehensive end-to-end evaluations with SPOT users to assess all components of the system, including the overall user experience.}

\section*{Acknowledgments}
This project is led by the Deutsche Welle Research and Cooperation Projects \update{team} and was co-funded by BKM ("Beauftragte der Bundesregierung für Kultur und Medien," the German Government's Commissioner for Culture and Media).



\section*{Limitations}
\update{While our approach performs well in several cases, it does not fully capture the complexity of real-world user queries. Users may phrase their queries ambiguously or use implicit descriptions rather than naming entities directly ('somewhere to eat' instead of 'restaurant', for example). In addition, references to entities with multiple interpretations, such as ambiguous landmarks, can introduce challenges that our current setup does not explicitly address. Another limitation is our reliance on OSM as the primary knowledge source. While OSM provides broad coverage, its data may be incomplete or inconsistent in certain regions. Addressing more diverse data sources and improving the handling of ambiguous or underspecified queries are important areas for future work.}

\section*{Ethics Statement}
SPOT \update{democratizes} access to geospatial data, but there are several ethical considerations. First, the underlying LLMs may contain inherent biases that could influence query interpretation and results. In addition, the OSM data itself has uneven coverage across regions, potentially limiting the utility of SPOT in under-represented areas.

Regional differences in tagging conventions also present challenges. Although our bundling approach mitigates some inconsistencies, cultural and regional idiosyncrasies in describing places may not be fully captured in our current implementation, reflecting potential limitations in the geographic perspective of the development team.

The most important ethical consideration is privacy. By lowering the technical barriers to geolocation identification, SPOT could potentially facilitate invasions of privacy through the analysis of images or videos shared on for example social media. While these capabilities already exist through tools such as Overpass Turbo, SPOT's accessibility heightens concerns. We believe that the benefits for legitimate fact-checking and investigative journalism outweigh these risks, but \update{emphasize} that users should only use SPOT for ethical purposes, such as verifying public information rather than tracking individuals. Ongoing work includes exploring additional safeguards to prevent misuse while preserving functionality for legitimate uses.

The broader impact of the tool lies in its potential to empower journalists around the world to verify information more efficiently, potentially countering misinformation and strengthening factual reporting in an era of increasing \update{manipulation of digital information}.

\bibliography{custom}

\clearpage

\appendix

\section{Appendix} 
\subsection{Requirements for SPOT} \label{appendix:requirements}
We list the requirements of SPOT that serve as also function for the data generation pipeline. 

\noindent
\textbf{Entity Recognition}
\begin{itemize}
    \item SPOT identifies general categories like restaurant, train station which allows recognition of places based on category type.
    \item SPOT detects specific brand names, including `McDonald's', `KFC', `Tchibo' and compound names such as Thalia bookstore.
\end{itemize}

\noindent
\textbf{Entity Properties}
\begin{itemize}
    \item SPOT  identifies properties such as `organic (food shop)', `Italian (restaurant)' or colors such as `brown (bench)' for refined queries. 
    \item SPOT interprets quantitative descriptors such as height, floors and house numbers.
\end{itemize}

\noindent
\textbf{Area Recognition}
\begin{itemize}
    \item SPOT supports cities, districts, and regions, including multi-word areas (e.g., "New York") and states such as "Nordrhein-Westfalen." 
    \item SPOT introduces bounding box support for identifying entities within a broader, undefined area. 
\end{itemize}

\noindent
\textbf{Relation Recognition}
\begin{itemize}
    \item SPOT interprets both numeric distances (e.g., `100 meters') and written forms (e.g., `one hundred meters'). 
    \item SPOT supports terms like `next to', `opposite from' and `beside' to improve natural understanding of spatial relationships. 
    \item SPOT supports distance-based relations 1) radius constraints (e.g. entity A to entity B and entity C) and entity chains (e.g. entity A to B and entity B to entity C).
    \item SPOT recognizes relationship such as `a fountain within a park' and `a shop inside a mall', `a park with a fountain', `hotel with a parking lot'.
\end{itemize}

\needspace{6\baselineskip}

\noindent
\textbf{Robustness to Different Styles}
\begin{itemize}
    \item SPOT can match descriptors with slight variations such as plurals ("bookshops" vs. "bookshop") and minor differences (`bookstore' vs. `book shop'). 
    \item SPOT is robust to typos in names and common words (e.g., `MacDonalds' for `McDonald's')
    \item SPOT is robust to styles that presents different user profiles such as an experienced fact-checker, beginner, etc. Additionally, it is robust to formal and casual query styles.
    \item SPOT recognizes area names and locations in code-switching texts (mixture of texts in different languages). For example, area and brand names in non-Roman alphabets such as Cyrillic and Greek.
    \item SPOT supports both single and multi-sentence structures in user queries.
\end{itemize}

\subsection{Relative Spatial Terms\label{appendix:relspat}}

\begin{table*}[ht!]
\small
    \centering
    \begin{tabular}{c c l}
        \toprule
        Index & Distance & Terms \\
        \midrule
        0  & 25 m   & not far away, enclosed by \\
        1  & 50 m   & next to, among, adjacent, beside, side by side, at, next door \\
        2  & 100 m  & near, around it, in close distance to, surrounded from \\
        3  & 150 m  & in front of, close to, opposite from, in surroundings \\
        4  & 250 m  & on the opposite side \\
        5  & 1000 m & on the edge \\
        6  & 2000 m & nearby \\
        \bottomrule
    \end{tabular}
    \caption{List of relative spatial terms and distance values used during data generation.}
    \label{tab:distance_descriptions}
\end{table*}
A list of relative spatial terms and their interpretation can be found in Table~\ref{tab:distance_descriptions}.

\subsection{Styles and Personas}\label{appendix:persona_style}
\update{Writing styles randomly selected in each prompt: ``in perfect grammar and clear wording'', ``in simple language'', ``with very precise wording, short, to the point'', ``with very elaborate wording'',``as a chain of thoughts split into multiple sentences'. }

\update{Personas randomly selected in each prompt: ``political journalist'', ``investigative journalist'', ``expert fact checker'', ``hobby fact checker'', ``human rights abuse monitoring OSINT Expert'', ``OSINT beginner'', ``legal professional''.}

\subsection{Prompts\label{appendix:prompting}}
\subsubsection{Dataset Generation}
We designed a dynamic prompt with some randomly selected parameters. 

\begin{table*}[!ht]
\small
  \adjustbox{width=\linewidth}{
\begin{tabular}{|p{0.2\linewidth}|p{0.65\linewidth}|p{0.15\linewidth}|}
        \hline
        \textbf{Tag Combination} & \textbf{Prompt} & \textbf{Generated Sentence} \\ \hline
\begin{lstlisting}[basicstyle=\ttfamily\tiny]
area:
  type: bbox
entities:
- id: 0
  name: church
  properties:
  - name: levels
    operator: '>'
    value: '56'
  type: nwr
- id: 1
  name: bridge
  properties:
  - name: name
    operator: '~'
    value: MK6
  type: nwr
relations:
- source: 0
  target: 1
  type: distance
  value: 16460 m
\end{lstlisting}
 &  
Generate one or more sentences simulating a user using a natural language interface for an AI geolocation search tool that finds locations based on descriptions of objects and their spatial relations. Each object has one main descriptor and optionally additional properties. All properties must be put in a logical connection to the object. Objects can either be single instances, or clusters of multiple of one object which are located in a specific distance radius (e.g. "three houses next to/within 10m of each other").
Mention the area, cover all entities and their respective properties, and describe the respective relations. Stick to the descriptions of entities and relations provided and don’t add anything. When describing names or brand (names), be creative in your phrasing (examples being a "book store of brand Thalia" vs. "a Thalia book store", or simply e.g. "a Thalia" if the type of object is not given). Stick to the values of each relation. Distances always refer to a maximum distance. If no distance is given, do not use any terms such as close, near, create sentences such as "find a house and a restaurant". Vary your phrasing. Do not affirm this request and return nothing but the answer.
 ==Persona==
hobby fact checker 
==Style==
as a chain of thoughts split into multiple sentences
==Input==
Objects:
- Obj. 0: church | Properties -> levels: above 56
- Obj. 1: bridge | Properties -> name:  "MK6"
Distances:
- All objects are no more than 16460 meters from another.
Please take your time and make sure that all the provided information is contained in the sentence.
& Looking around an area, I'm trying to find a church that has more than 56 levels. In the same vicinity, not exceeding a distance of 16,460 meters, there should also be a bridge called "MK6".\\
\bottomrule
    \end{tabular}}
    \caption{An example parametric prompt used for data generation. Due to space limitations, the prompt formatting was altered. The original prompts can be found in the source code.}
    \label{tab:app22}
\end{table*}

An example of the generated sample is shown in Table~\ref{tab:app22}.

\subsubsection{Inferencing Prompt}

\begin{figure*}[!ht]
\tiny
    \centering
    \begin{tcolorbox}[colback=blue!5!white, colframe=blue!75!black, title=Inferencing Prompt, label=box:opinion_analysis]
You are a joint entity and relation extractor. Given a text that is provided by geo fact-checkers or investigative journalists, execute the following tasks:

1. Identify the area mentioned in the text. If no area is found, designate its type as 'bbox' and assign its name as 'bbox'. If area is found, designate its type as 'area'.

2. Detect and extract the geographical entities present in the text. Areas are not part of these entities. Entities are always present in a sentence. There are two type of entities: cluster and nwr. The 'cluster' type is clusters of entities, allowing queries like "3 Italian restaurants next to each other" or "at least 5 wind generators nearby." The other entity types belongs to nwr.

3. Extract properties associated with each identified entity, if available. The properties must be related to their types, colors, heights, etc.

4. Identify and extract any relations between the entities if mentioned in the text. We define two relation types: contains and dist. Assign one of them as the relation type. In contains relations, you can recognize relationships such as "a fountain within a park" and "a shop inside a mall.". In contain relation, there is no distance. In dist relation, you interpret both numeric distances (e.g., "100 meters") and written forms (e.g., "one hundred meters"), support terms like "next to," "opposite from," and "beside" to improve natural understanding of spatial relationships, and recognize Multiple distance-based relations are supported, including radius constraints (e "A to B and C") and entity chains (e.g., "A to B and B to C").

Let's think step by step.

Please provide the output as the following YAML format and don't provide any explanation nor note:

\begin{verbatim}
area:
   type: area type
   value: area name
entities:
 - name: [entity name 1]
   id: [entity id 1]
   type: [entity type 1]
   properties:
    - name: [property name 1]
      operator: [operator 1]
      value: [property value 1]
    - name: [property name 2]
      operator: [operator 2]
      value: [property value 2]
    - ...
 - name: entity name 2
   id: entity id 2
   type: entity type 2
 - ...
relations:
 - source: entity id 1
   target: entity id 2
   type: relation between entity 1 and entity 2
   value: relation distance if the type of relation is dist
 - ...
\end{verbatim}
    \end{tcolorbox}
    \caption{Zero-shot prompt used to query the LLMs, containing instructions and the YAML layout. The prompt includes support for cluster-type entities, which were not available in the deployed system at the time of writing.}
    \label{fig:zero_shot}
\end{figure*}

For the one-shot prompt, we appended one sample from the training data to the zero-shot prompt. The matching of each benchmarking samples to one training sample is based on the cosine similarity of their SBERT embeddings.

\end{document}